\documentstyle[12pt]{article}
\def \F{\phi}

\def\NP{{\it Nucl. Phys.\ }}

\def\PL{{\it Phys. Lett.\ }}
\def\PR{{\it Phys. Rev.\ }}

\def\IJMP{{\it Int. Jour. Mod. Phys.\ }}

\def\l{\lambda}

\def\s{\sigma}

\def\half{{1\over 2}}
\def\d{\dagger}
\def\be{\begin{equation}}
\def\eq{\end{equation}}
\def\Tr{{\rm Tr}}

\begin{document}

\begin{flushright}
OUTP-9430P\\
hep-th/9411204\\
\end{flushright}
\vspace{20mm}
\begin{center}
{\LARGE  $c>1$ Non-Critical Strings  and Large-$N$ \\
Matrix Field Theory\\}
\vspace{30mm}
{\bf F.Antonuccio and S.Dalley}\\
\vspace{5mm}
{\em Department of Physics, Theoretical Physics\\
1 Keble Road, Oxford OX1 3NP, U.K.}\\
\end{center}
\vspace{30mm}

\abstract
Motivated by a possible relativistic string description of hadrons we use a
discretised light-cone quantisation and Lanczos algorithm to investigate the
phase structure of $\F^3$ matrix field theory in the large $N$ limit. In $1+1$
dimensions we confirm the existence of Polyakov's non-critical string theory
at the boundary between parton-like and string-like phases, finding critical
exponents for longitudinal oscillations equal to or consistent with those
given by a mean field argument. The excitation spectrum is finite, possibly
discrete. We calculate
light-cone structure functions and find evidence that the probability
$Q(x)$ of a parton in the string carrying longitudinal momentum
fraction
between $x$ and $x+dx$
has support on all $0<x<1$, despite the average number of partons being
infinite.

\newpage
\baselineskip .25in
\section{Light-Cone Strings.}

Soon after the seminal work of Goddard et al. \cite{god} on
canonical
light-cone quantisation of relativistic string theory, it was realised
that the restriction to critical central charge $c$  of matter
might be avoided if
motions of the string were allowed which made it impossible to go to
light-cone gauge. Point-like insertions of energy-momentum
\cite{pat,gre1} result in a singular space-time
co-ordinate function
$X(\s )$ so that a smooth change of
variable
$\s \to \s'$ on the string
is unable to redistribute the longitudinal momentum $P^+$
evenly along the string. In this way the longitudinal oscillations
of the string remain as physical degrees of freedom.
However the longitudinal oscillations
are  in general  non-linear and the difficulty in studying such systems
has
left a dearth of reliable results for
non-critical strings in physical spacetime dimensions. These
theories are  interesting as models for hadrons,
this being the principal motivation behind the work
of Green \cite{gre1} for example. Point-like insertions of energy momentum
on the string give it a character intermediate between that of  critical
strings and a finite collection of partons, and therefore there is a
possibility to describe both the parton-like and string-like behaviour
of hadrons in a single transparent framework.
Matrix field theories \cite{wei,dav} offer an explicit implementation
of this idea.
In this paper we continue the study of the discretised
light-cone quantisation (DLCQ) of
matrix models with $c \geq 2$ introduced in ref.\cite{dal2} and
further studied numerically \cite{kres} and analytically \cite{dal1},
by applying a Lanczos algorithm to the mass-matrix evaluation.
We find a parton-like phase, and also  string-like phase
in which
longitudinal dynamics are essentially frozen, corresponding to
the string of Goddard et al. at non-critical $c$ \cite{tho1,igo2}.
These two phases are
separated by a transition at which there exists a non-critical string
theory, essentially that of Polyakov \cite{pol},
with non-trivial longitudinal degrees of freedom. Although a
stringy object it seems to exhibit parton-like
behaviour in structure functions.

We consider an $N$x$N$ hermitian matrix field ${\F}_{ab}(x)$ in $d$ dimensions
subject to the following (Euclidean) action
\be
 S=\int d^d x \Tr \left( \half (\partial_\alpha \F)^2+\half \mu \F^2-
{1\over 3\sqrt{N}}\lambda \F^3 \right) \ .\label{action}
\eq
It is well-known that the Feynman diagram $1/N$ expansion, when considered
in terms of the dynamical triangulations given by the dual graphs
\cite{dav,jan1},
represents an explicit discretisation of the Polyakov path integral
\cite{pol}
for random surfaces, albeit with an exponential rather than Gaussian
propagator. It may be regarded as a prototype
QCD in the sense that it is a theory of fields transforming in the
 adjoint representation of (in this case global) $SU(N)$, from which
one can build a flux-tube/string representation of the bound states.
In the context
of light-cone quantisation the relationship is more precisely stated
in ref.\cite{dal3} as a dimensional reduction.
The observables we consider are the closed strings of $n$
partons at some
fixed
time (prototype glueballs) ${\rm Tr} [ \F (x_{1}) \ldots \F (x_{n}) ] $ and
products of these (multi-string states).
If mass eigenstates have $<n>$ finite they
will be
called
parton-like, while if  $<n> = \infty$ they will be referred to as
string-like.
Applying light-cone
quantisation
in Minkowski space, one derives the spectrum as a function of
the bare worldsheet cosmological constant $\log{\lambda}$
and can search for critical behaviour in this parameter.
Such behaviour can occur since, according to
general arguments \cite{div}, the sum of graphs at a given order
in $1/N$ in a UV
finite theory should itself be finite for sufficiently
small coupling constant $\l$; the
$1/N$ expansion on the other hand is only asymptotic as a result of the
unboundedness of the
action (\ref{action}). As $\lambda \to \lambda_{c}$ one
approaches the edge of the domain of convergence, at which large
graphs
are favoured, which in light-cone formalism means a transition from a
parton-like to a string-like phase. We shall consider the $N \to
\infty$
limit and study the UV-finite normal-ordered
$1+1$-dimensional theory,
commenting at the end on the more
general
case.

Let us now rotate $x^{0} \to ix^{0}$ and find the relativistic
spectrum by light-cone quantisation, treating $x^{+}=(x^0 + x^1)/\sqrt
2$ as time and $k^{+} = (k^0 + k^1)/\sqrt{2} > 0$ as (longitudinal)
momentum.  This has been described before
\cite{dal2} and so we make only brief comments.
The problem can be expressed in terms of the Fourier modes
$a_{ij}(k^+)$ of $\F_{ij}$. We will
discretize the problem to render the number of
states of total momentum $P^+$ finite by allowing partons with
$ k^+ = m P^+ /K$ where $m$ is a positive integer
and $K \geq m$  is a  fixed large integer which plays the
role of cutoff \cite{tho1,brod}. Then in the quantum theory
\be\delta_{mm'} \delta_{il} \delta_{jk}  =
[a_{ij}(m),a_{lk}^{\d}(m')] \ ,
\eq
and
\be
2P^{+}P^{-}  = \mu K (V -y T) \ , \ \ y={\lambda \over 2\mu \sqrt{\pi}}
\ ,\label{ham}
\eq
where
\begin{eqnarray}
V & =& \sum_{m=1}^{K} {1\over m}  \ a_{ij}^{\d}(m) a_{ij}(m) \ ,
\label{disc1}
\\
T & =& {1 \over \sqrt{N}} \sum_{m_1 , m_2 =1}^{K}  {a_{ij}^{\d}(m_1
+m_2)a_{ik}(m_2)a_{kj}(m_1)
+ a_{ik}^{\d}(m_1)a_{kj}^{\d}(m_2)a_{ij}(m_1 +m_{2}) \over
\sqrt{m_{1}m_{2}(m_{1} + m_{2})}} \ . \label{disc2}
\end{eqnarray}
Single closed-string states are in general
linear combinations of the Fock states
\be
\Psi  = \sum_{n=1}^{\infty}
 \sum_{m_{i}=1}^{K} \delta (K - \sum_{i=1}^{n} m_i )
f_{n} (m_{1}, \ldots ,m_{n}) N^{-n/2}
{\rm Tr}[a^{\dagger}(m_{1})
\cdots a^{\dagger} (m_{n})] |0>\ .\label{super}
\eq
The light-cone energy $P^-$ satisfies
$:P^- : |0> = 0$.
The spectrum $M^2 = 2 P^+ P^-$  can be found by
diagonalising the matrix eigenvalue problem (\ref{ham}) in the basis of
discretised Fock states, labelled by the ordered partitions of
$K$.\footnote{A
 simple example at
$K=3$ can be found in ref.\cite{dal2}.}
To establish the basic stringy scaling laws we will study  the $N \to
\infty$
free-string limit (not equivalent to free partons!). There are many
fewer states to consider in this case, since $1/N$ is the string
coupling and only single string
states need be retained,  leading to a linear
light-cone Schrodinger equation for $\Psi$ \cite{tho1,igo2}.
The eigenfunctions  give directly the structure functions of the string in
terms of the longitudinal momentum fractions $x= m/K$.
The light-cone quantisation we performed is naive in the sense that we
have neglected zero modes  $a^\dagger (0)$. This was done partly for
simplicity but also to regulate the theory in the string phase as will
become
clear shortly.

\section{Numerical Results}

The number of Fock states increases exponentially with $K$
and the exact determination (and diagonalisation) of the matrix
 representation of $M^2$ soon becomes onerous, e.g. at
$K=13,14,15,16$ there are $631,1181,2191,4115$ states respectively.
Nevertheless, if we wish to consider only the extreme eigenstates
of an operator, such as the ground state and near ground states, then
a considerable amount of computer processing time can be saved
by implementing numerical approximation schemes that do not require
a precise knowledge of the operator's matrix representation.
We used the Lanczos algorithm,
in the context of DLCQ specific applications of which are detailed
in ref.\cite{hill}. In the context
of our own work, all the necessary numerical and symbolic
manipulations were performed with the help of Mathematica and  Sun
workstations.
The Lanczos algorithm is iterative by nature, and a judicious
choice of initial state  is necessary to ensure
proper  convergence.
It was soon discovered that any straightforward application
of this method involves the manipulation of a rapidly growing number
of states  at each subsequent step in the iteration.
This arises because successive application of $P^-$
on a typical Fock state  yields a proliferation of
new states that grows rapidly as the iteration proceeds, and it is
this property that severely limits the scope of any iterative
approximation scheme.
Progress can be made, however, by recognizing that the action of
the hamiltonian on existing Fock states
is repeated many times during the iteration process,
so unnecessary computations can be avoided by `remembering'
where a single string Fock state ends up after it has first been acted
on by the light-cone hamiltonian. In computing terminology,
this amounts to introducing `pointers' between Fock states
during the course of the programme.
Implementation of this scheme enabled $K \leq 16$ to be studied.
The one parton state $N^{-1/2} \Tr[a^{\d}(K)]|0>$ seems an adequate
initial state for
the Lanczos algorithm, and convergence was assessed to at least 3 s.f.
accuracy in anything we plot in this paper.

In ref.\cite{dal1} a mean field theory was used to map out the
structure
of the lowest mass eigenstate of the
$1+1$-dimensional theory (\ref{action}) at $K=\infty$ as $y =\l / 2 \mu
\sqrt{\pi}$
is increased.  At $y=0$ this is the one-parton state.
 For $y < y_c \approx 0.53$
states with a few partons dominate in the
wavefunction, while at $y > y_c$  each parton
typically carries a finite amount of discretised momentum $m$, which
decreases with increasing $y$, and the wavefunction typically carries
an infinite number of partons. In this latter phase the longitudinal dynamics
are trivial, essentially those of a theory with all $m=1$ ($n =K$)
\cite{tho1,igo2}. This picture is only valid when zero modes are
neglected -- for a
given
$K$ this regulates the instability to creation of more and more
partons
when $y > y_c$.
At $y = y_c$ on the other hand
there are typically an infinite number of partons each
carrying an infinite amount of discrete momentum $m$, so that
the longitudinal dynamics are non-trivial.
At $y > y_c$ mean field predicted
the
scaling laws $<n> \propto K$ and $M^2 \propto  -K^2$ as $K \to \infty$, laws
which appear to be exact, and so the quantities $M^2/K^2$ and $<n>/K$
can be used as a signal for the string phase.
We plot them in Figs. 1(a) and 1(b),
both for the mean field\footnote{There is a scale ambiguity in the
mean field coupling $y$ which we have fixed from the position of the
true $y_c$ estimated from the exact diagonalisations.}
  ($K= \infty$) and an exact
diagonalisation
of the matrix $M^2$ at $K=12$ ($351$ states). In fact mean field
predicts that as $y - y_c \to 0^+$, $ M^2 /K^2 \propto -(y-y_c )^2$ and
$<n>/K \propto (y-y_c )$.

The probability $P(n)$ of finding $n$ partons and the probability
density $Q(x)$ of
finding a parton with momentum fraction $x$ in the groundstate gives
information about the internal structure of the string in each phase.
We computed these functions for $y < y_c, y \approx y_c$, and $y>y_c$.
At $y<y_c$ we typically found that $P(n)$ and $Q(x)$ were
essentially delta-functions
at $n=1$ and $x=1$, indicating that very little mixing into higher
Fock states occurs. In Figs. 2(a) and 2(b) we plot them for $y = 0.53 $
and $K=12$.
The length distribution of strings $P(n)$ is still peaked at $n=1$, a
peak
which falls with increasing $K$, but
now has contributions from all other lengths. Similarly the structure
function
$Q(x)$ is still peaked at $x=1$ but with a shallow secondary peak at
$x \sim 0.5$, the latter presumably a manifestation of the dominant
$1 \to 2$ decay mode of the one parton state --- mean field predicts that
$<n> \sim  2$ at $K=12$ and $y=y_c$.
For $y > y_c$ the behaviour is quite different, shown in also at
$K=12$ in
Figs. 2(a) and 2(b) for $y=1$. Now $P(n)$ is peaked at $n \propto K$ with
increasing $K$, while $Q(x)$ is peaked
at low $x$ already at $K=12$. This indicates that the partons carry a finite
amount of discrete momentum in this phase, while at the transition point
this is was not necessarily true. Figure 3 shows how $Q(0.5)$ at $y=0.53$
changes
with $K$. Remarkably it seems to tend to a non-zero value as $K \to
\infty$,
indicating that although $<n> =\infty$ and we are dealing with a
string
theory, the structure function has finite support on the whole
interval
$(0,1]$. We checked also that at $y > y_c$, $Q(0.5)$ falls with
increasing
$K$ and is consistent with zero at $K = \infty$.

To estimate the critical exponents $<n> \propto K^{\gamma}$ and
$M^2 \propto K^{\delta}$ we used  log-log plots with increasing $K$.
For $y > 0.53$ we found rapid convergence already  at small $K$,
showing with certainty that the mean field exponents are exact:
$\gamma = 1$; $ \delta =2$.
Similarly rapid convergence both as a function of $K$ and number
of iterations of the Lanczos algorithm is found in the parton
phase $y<y_c$. For $y \approx y_c$ the large fluctuations
charateristic
of a critical point make convergence much slower. We expect the forms
\begin{eqnarray}
{<n> \over K} &  = &a_1 (y-y_c )^a + a_2 K^{\gamma -1} + \cdots \\
{M^2 \over K^2} &= & b_1 (y-y_c )^b + b_2 K^{\delta -2} + \cdots \label{exp}
\end{eqnarray}
where the ellipses indicate higher orders in $1/K$. Mean field
predicts
$a=1$ and $b=2$, which we assume are not far from the true values (see
fig.1).
If
we
also assume that the coefficients $a_i$ and $b_i$ are not too large
or small then
$\gamma$ and $\delta$ can be estimated provided we get close enough
to the critical point and have high enough $K$. Figure 4 plots
$\log{[<n>/\sqrt{K}]}$ at $y=0.53$.
We are confident that the true $y_c$ is
within $0.01$ so that with the above assumptions finite $K$ is
really the only limitation since we typically measured
$<n>/K \sim 0.1$. There is a  degree
of uncertainty since we have  not clearly
reached a straight-line scaling region, which would be horizontal if
the mean field exponent were exact, but taking the slope at
the highest $K$ gives an upper bound
at $y = y_c$ of $ \gamma < 0.6 \ ({\rm meanfield} \ 0.5)$.
At $K \sim 16$ we find that  $M^2 / K^2 \sim - 0.0001$ at $y=0.53$,
which is still probably
of
the same order as the leading term in eq.(\ref{exp}). Therefore
we
were unable to estimate $\delta$ in this case (mean field value
$\delta = 1$).

For $y > y_c$ we have found similar
scaling
laws hold for excited states, whose squared masses are separated
by $O(K^2)$ gaps. Thus the longitudinal {\em fluctuations} with
respect to the groundstate are indeed trivial in this phase.
The behaviour of excited states with respect to the groundstate
was also studied as a function of $y$ in refs.\cite{dal2,kres}, with
the
result that the gaps in the spectrum appeared finite for $y < y_c $.
To study this question
as $y - y_c \to 0^-$ we calculated the mass difference of ground and
first excited state at $y=0.53$ for increasing $K$, shown in Figure 5.
It clearly favours a finite spectrum. Unfortunately the data
cannot decide for sure between continuous or discrete spectrum, but
the latter is not ruled out.

\section{Discussion}

In the limited space left to us here let us address a couple of the
questions posed by the preceeding analysis.
We have analysed the $1+1$ dimensional theory explicitly here, but in
principle one could add some transverse degrees of freedom. In order
to assess the effect of these we studied the simplest case, that
 of Ising degrees of
freedom  on the string, i.e. a transverse lattice consisting of two
points.
This can be solved in a mean field approximation \cite{dal5} by an analysis
similar to that in ref.\cite{dal1}, and we found that the exponent
$\gamma$
was unchanged for all values of the Ising model coupling constant. If
the
mean field is trustworthy, it leads one to believe that transverse
degrees
of freedom do not change the scaling laws for the longitudinal modes.
It is of course quite
possible
that the excitation spectrum acquires additional states at finite
energy
with respect to the groundstate for some critical Ising coupling, a
fact
which can be proved explicitly \cite{dal5} in the $y \to \infty$ limit
along the lines
of
refs.\cite{tho1,igo2,dal4}.

Although the $1/N$ expansion of the field theory (\ref{action}) has no
pathologies as such, we have taken various steps to isolate a string
theory, such as neglecting non-singlets and zero modes, which
questions the
self-consistency. A full discussion of consistency is beyond the scope
of this
letter but we mention the problem of renormalisation of the mass
divergence. Presumably this can be cured by some procedure analogous
to
those employed for critical strings, where there is a similar
divergence
(of opposite sign howvever). For example, mean
field predicts that the groundstate mass squared diverges like
$\sim -K$ at $y=y_c$ which, if correct,
could be subtracted by adding a constant to the
Lagrangian density
(\ref{action}). Further consistency conditions may
serve to fix the finite part, but we see no precise reason at this
stage
why it
should be
tachyonic.

Our results have suggested
that the free non-critical bosonic string coupled to a
point source   should exhibit a parton-like structure
function, and we could not rule out discrete non-tachyonic spectrum.
Moreover the work of Green \cite{gre1} indicated that high energy fixed
angle
scattering of strings similar to ours shows power law behaviour.
All these properties are of course those of actual hadrons.
Obviously our numerical  results must be
investigated further to check that they are not  finite cutoff artifacts,
but
there seems no obstacle to more extensive numerical calculations using
the
Lanczos algorithm we have implemented here, and it seems worthwhile to
pursue with renewed vigour analytic approaches to  $c>1$ non-critical
string theory.

\vspace{10mm}
\noindent Acknowledgements: We thank I.Klebanov for helpful
interactions. F.A. is supported by the Commonwealth Scholarship and
Fellowship Plan (British Council).

\vfil

\newpage
\begin{center}
FIGURE CAPTIONS
\end{center}
\noindent Fig.1. -- (a) $M^2/K^2$ for the groundstate ($\mu =1$).
(b) $<n>/K$ in the groundstate. Broken line
is mean field result at $K=\infty$, solid line is exact result at
$K=12$.

\noindent Fig.2. -- (a) Probability $P(n)$ of finding $n$ partons in
the
groundstate. (b) Probability $Q(x)$ of finding a parton with
longitudinal
momentum fraction between $x$ and $x + dx$ in the groundstate. Solid
circles
are $y=0.53$, open circles are $y=1$.

\noindent Fig.3. -- $Q(0.5)$ for increasing cut-off $K$ at $y=0.53$

\noindent Fig.4. -- $\log{<n>/\sqrt{K}}$ for increasing $K$ at
$y=0.53$.

\noindent Fig.5. -- $\Delta M^2 = M_{1}^2 -M_{0}^2$, the difference
in mass
squared $(\mu =1)$
between ground and first excited state, for increasing $K$ at
$y=0.53$.
\vfil
\end{document}